\documentclass[aps, reprint, amsmath,amssymb,]{revtex4-1}

\usepackage{dcolumn}
\usepackage{graphicx,latexsym,amsfonts,bm}

\begin{document}
\preprint{AIP/123-QED}

\title{Diffusive instability of a Townsend discharge}
\author{V.V. Mikhailenko}
\altaffiliation{Pusan National University, 30 Jangjeon-dong, Guemjeong-gu, Pusan 609--735, S. Korea.}
  \email{vladimir@pusan.ac.kr}
\author{H.J. Lee}%
\affiliation{Pusan National University, 30 Jangjeon-dong, Guemjeong-gu, Pusan 609--735, S. Korea.}
\author{V.S. Mikhailenko}
\affiliation{V.N. Karazin Kharkov National University, 61108 Kharkov, Ukraine.}

\date{\today}

\begin{abstract}
The role of the electron diffusion on the stability of a Townsend discharge is investigated. It is obtained, that electron
 diffusion modifies the condition of the steady self-sustenance of the discharge, and make discharge unstable.
\end{abstract}


\maketitle

\section{Introduction}

There has been considerable interest in non-thermal atmospheric pressure glow discharges over last time due to the increased variety of their industrial applications\citep{Fridman,Lee}. The basic feature of the non-thermal discharges is that majority of the energy of the applied electric field goes into electrons, instead of heating the entire gas in the discharge cell. Numerous experiments show that depending on the parameters of the discharge, atmospheric pressure glow discharge is realized in two forms: a Townsend and glow discharges. A Townsend discharge is the simplest type of glow discharge. It is characterized by the absence of quasi-neutral plasma -- the absolute value of the ion density exceeds much that of the electron density. The applied electric field is weakly disturbed by spatial charge and the discharge current is governed mainly by the processes of the electron emission from the cathode. The current of a Townsend discharge is only limited by the external circuit and when the space charge in a Townsend discharge becomes large enough to cause a significant disturbance of the applied field, the transition to glow discharge occurs.

The physics of space charge driven transitions of a Townsend discharge to subnormal, normal and further to abnormal glow has drown considerable attention of the discharge investigating community (Refs.\citep{Melekin}--\citep{Amiranashvili} and references therein). Numerous experimental, analytical and numerical investigations provide deep insight into amazing variety of spatio-temporal processes, which are responsible for such transitions. In this paper, the stability of a Townsend discharge is investigated for the unexplored yet regimes far from such transitions, when space charge is too small to produce any significant distortion of the applied electric field. We find, however that even in that case narrow planar Townsend discharge, with the distance between the electrodes considerably smaller than the radius of the discharge cell, appears unstable. The discovered instability obtains analytical confirmation as a resulted from electron diffusion in the axial direction. The derivation of the basic equations and boundary conditions for the dimensionless variables corresponding to the regime of the Townsend discharge is presented in Section 2. In Section 3, we present the results of the calculations of the modified by electron diffusion condition for the steady self-sustenance of the Townsend discharge and investigate its stability under that condition. A summary and discussion is presented in Sec.4.  

\section{Basic equations and boundary conditions}
The simplest set of equations containing the basic physics necessary for the investigation of the glow discharge stability comprises the well-known continuity equations in the drift-diffusion approximation for electrons $N_{e}$ and positive ions $N_{i}$, coupled with Poison equation for the electrostatic potential $\Phi$ (e.g.,\cite{Kolobov}),
\begin{eqnarray}
&\displaystyle\frac{\partial N_{i}}{\partial t}-\frac{\partial }{\partial Z}\mu _{i}N_{i}E =N_{e}\alpha\left(E\right)\mu _{e}E,  \label{1} \\
&\displaystyle\frac{\partial N_{e}}{\partial t}+\frac{\partial}{\partial Z} \left[
-D_{e}\frac{\partial N_{e}}{\partial Z} +\mu_{e}N_{e}E \right] =N_{e}\alpha\left(E\right)\mu _{e}E,  \label{2} \\
&\displaystyle\varepsilon_{0}\frac{\partial E}{\partial Z} =e\left(
N_{e}-N_{i}\right).  \label{3}
\end{eqnarray}
Here  $D_{e}$, $\mu_{i}$, and $\mu_{e}$ are the electron
diffusion coefficient and mobilities of ions and
electrons, respectively; $\alpha\left(E\right)$ is Townsend's ionization coefficient (e.g.,
\cite{Raizer}) and $E$ is  the electric field; $\varepsilon_{0}$ is the
permittivity of free space, $e$ is the elementary charge, the $Z$-axis is
directed from the cathode to the anode, and $t$ is time. For considered here Townsend limit, both mobility and electron diffusion coefficient, which otherwise depend on the local electric field, can be assumed as constant.
Because in the Townsend mode electric field is practically constant, local field approximation is applicable to ionization rate $\alpha$. Farther, the process of dissociative recombination and ion diffusion (we assume that electron temperature
exceeds greatly the ion temperature) are completely neglected. In contrast, the axial diffusion of
electrons is taken into account.
The boundary conditions are taken in the Townsend form. At the cathode $(Z= 0)$
\begin{eqnarray}
&\displaystyle -D_{e}\frac{\partial N_{e}}{\partial Z}
+N_{e}\mu _{e}E = \gamma N_{i}\mu_{i}E,\label{4}
\end{eqnarray}
where $\gamma$ is the secondary electron emission coefficient and electrons flax from cathode includes
mobility and diffusive flaxes. At the anode $(Z = L)$:
\begin{eqnarray}
&\displaystyle \frac{\partial N_{e}}{\partial Z} = 0,\quad N_{i} = 0, \label{5}
\end{eqnarray}
and the electric current density at the anode is given by
\begin{equation}
j=-eN_{e}\mu _{e}E+\varepsilon_{0}\frac{\partial E}{\partial t}. \label{6}
\end{equation}
The boundary conditions at the wall of the discharge vessel are
not relevant for the present one-dimensional study.

In what follows we are interested in processes that are assumed to be on time scale which is
longer than the ion travel time
\begin{eqnarray}
&\displaystyle t_{0}=\frac{L}{\mu_{i}E}, \label{7}
\end{eqnarray}
and therefore ion density time derivative will be considered as a small perturbation.
On that time scale the electron density time dependence is eliminated adiabatically and electron density time derivative in Eq.(\ref{2}) will be setting to zero. It is convenient to introduce the dimensionless times $\tau$, length $z$ and electric field $\hat{E}$,
\begin{equation}
\tau=\frac{t}{t_{0}}, \quad z=\frac{Z}{L}, \quad \hat{E}=\frac{E}{E_{t}}\label{8}
\end{equation}
and parameter $\varepsilon_{e}$,
\begin{equation}
\varepsilon_{e}=\frac{D_{e}}{\mu_{e}E_{t}L}=\frac{T_{e}}{eLE_{t}}, \label{9}
\end{equation}
which determines the relative value of the diffusive and drift terms in Eq.(\ref{2}). By assuming that dimensionless anode current is equal to unity, we obtain from boundary condition (\ref{9}) the following dimensionless electron density $n_{e}$:
\begin{equation}
n_{e}=\frac{N_{e}e\mu_{e}E_{t}}{j}.\label{10}
\end{equation}
Similar relation,
\begin{equation}
n_{i}=\frac{N_{i}e\mu_{i}E_{t}}{j}.\label{11}
\end{equation}
we use for the dimensionless ion density $n_{i}$. In dimensionless form, system (\ref{1})--(\ref{3}) is
\begin{equation}
\frac{\partial n_{i}}{\partial\tau}-\frac{\partial}{\partial z}\left(n_{i}\hat{E} \right)=\hat{\alpha}n_{e}\left|\hat{E}\right| .\label{12}
\end{equation}
\begin{equation}
-\varepsilon_{e}\frac{\partial^{2} n_{e}}{\partial z^{2}}+\frac{\partial}{\partial z}\left(n_{e}\hat{E} \right)=\hat{\alpha}n_{e}\left|\hat{E}\right| ,\label{13}
\end{equation}
\begin{equation}
\frac{\partial \hat{E}}{\partial z}=\delta\left( \frac{\mu_{i}}{\mu_{e}}n_{e}-n_{i}\right)\approx-\delta n_{i} ,\label{14}
\end{equation}
where $\hat{\alpha}=\alpha L$. Parameter $\delta$,
\begin{equation}
\delta=\frac{jL}{\varepsilon_{0}\mu_{i}E^{2}_{t}} ,\label{15}
\end{equation}
 which, as it follows from Eq.(\ref{14}), determines the measure of the distortion of the external ambient electric field by space charge\cite{Benilov}, as well as parameter $\varepsilon_{e}$, are assumed to be small for a Townsend discharge.

In dimensionless variables, boundary conditions (\ref{4})--(\ref{6}) are at the cathode $(z=0)$
\begin{equation}
-\varepsilon_{e}\frac{\partial n_{e}}{\partial z}+n_{e}\hat{E}=\gamma n_{i}\hat{E},\label{16}
\end{equation}
and at the anode $(z=1)$
\begin{equation}
n_{i}=0, \quad \quad \frac{\partial n_{e}}{\partial z}=0 ,\label{17}
\end{equation}
\begin{equation}
-1= n_{e}\hat{E}+\frac{1}{\delta}\frac{\partial \hat{E}}{\partial \tau}.\label{18}
\end{equation}
It is interesting to note, that derivative $\partial \hat{E}/\partial \tau$ enters only into condition (\ref{18}) at the anode, therefore the time dependence of electric field will be determined with boundary condition (\ref{18}). It follows from Eq.(\ref{14}), that
\begin{equation}
\hat{E}\left(z,\tau \right) =-\delta\int^{z}_{1}n_{i}\left( \zeta,\tau \right)d\zeta +\hat{E_{0}}+\delta \mathcal{E}\left(\tau \right). \label{19}
\end{equation}
In Eq.(\ref{19}) we have accounted for that the distortion $\mathcal{E}$ of the applied electric field $\hat{E_{0}}$ is resulted from space charge. It follows from Eq.(\ref{19}), that
\begin{equation}
\hat{E}\left(z=1,\tau \right)=\hat{E_{0}}+\delta \mathcal{E}\left(\tau \right). \label{20}
\end{equation}
With electric field (\ref{20}) boundary condition at the anode becomes
\begin{equation}
-1=-n_{e}\left(z=1,\tau \right)\left(1+\delta \mathcal{E}\left(\tau \right)\right)+\frac{d\mathcal{E}}{d\tau}, \label{21}
\end{equation}
where $\hat{E_{0}}=1 (E_{0}=E_{t})$ was used. The basic set of equations (\ref{12})--(\ref{14}) as well as boundary conditions does not contain a time in an explicit form. Therefore in our linear stability analysis  of the Townsend discharge we consider ion and electron densities in a conventional form
\begin{equation}
n_{e,i}\left(z,\tau; \varepsilon_{e},\delta\right)= n_{e0,i0}\left(z; \varepsilon_{e},\delta\right)+n_{e1,i1}\left(z; \varepsilon_{e},\delta\right)e^{\lambda\tau},\label{22}
\end{equation}
where $n_{e0,i0}$ is the equilibrium electron (ion) charge and $n_{e1,i1}$ are their time-dependent small perturbations. Using the expansions
\begin{equation}
n_{e0,i0}\left(z,\tau; \varepsilon_{e},\delta\right)= n^{(0)}_{e0,i0}\left(z; \varepsilon_{e}\right)+\delta n^{(1)}_{e0,i0}\left(z; \varepsilon_{e}\right),\label{23}
\end{equation}
\begin{equation}
n_{e1,i1}\left(z,\tau; \varepsilon_{e},\delta\right)= n^{(0)}_{e1,i1}\left(z; \varepsilon_{e}\right)+\delta n^{(1)}_{e1,i1}\left(z; \varepsilon_{e}\right),\label{24}
\end{equation}
in Eq.(\ref{21}) we obtain
\begin{eqnarray}
&\displaystyle
-1+n^{(0)}_{e0}\left(z=1; \varepsilon_{e} \right)+\delta n^{(1)}_{e0}\left(z=1; \varepsilon_{e} \right)
\nonumber \\
&\displaystyle=
\frac{d\mathcal{E}}{d\tau}-n^{(0)}_{e1}\left(z=1; \varepsilon_{e}\right)e^{\lambda\tau}-\delta n^{(1)}_{e1}\left(z=1; \varepsilon_{e}\right)e^{\lambda\tau}  \nonumber \\
&\displaystyle -\delta \mathcal{E}\left(\tau \right)\left(n^{(0)}_{e0}\left(z=1; \varepsilon_{e} \right)+ n^{(0)}_{e1}\left(z=1; \varepsilon_{e}\right)e^{\lambda\tau}\right)   \label{25}
\end{eqnarray}
For stationary electron density $n^{(0)}_{e0}$ , Eq.(\ref{25}) gives known boundary condition
\begin{equation}
n^{(0)}_{e0}\left(z=1; \varepsilon_{e}\right)=1,\label{26}
\end{equation}
as well as the equation for $\mathcal{E}$,
\begin{equation}
\frac{d\mathcal{E}}{d\tau}=n^{(0)}_{e1}\left(z=1; \varepsilon_{e}\right)e^{\lambda\tau},\label{27}
\end{equation}
which, for initial condition $\mathcal{E}\left(\tau\rightarrow -\infty \right)=0 $ has solution
\begin{equation}
\mathcal{E}\left( \tau\right)=\lambda^{-1}n^{(0)}_{e1}\left(z=1; \varepsilon_{e}\right)e^{\lambda\tau}.\label{28}
\end{equation}
It also follows from Eq.(\ref{25}) that
\begin{equation}
n^{(1)}_{e0}\left(z=1; \varepsilon_{e}\right)=0\label{29}
\end{equation}
and
\begin{eqnarray}
&\displaystyle
n^{(1)}_{e1}\left(z=1, \tau; \varepsilon_{e}\right)=-\mathcal{E}\left(\tau\right)=-n^{(0)}_{e1}\left(1; \varepsilon_{e}\right)\frac{e^{\lambda\tau}}{\lambda}.\label{30}
\end{eqnarray}
It stems from Eq.(\ref{30}), that  expansion (\ref{24}) is convergent for not small $|\lambda|$,
for which
\begin{equation}
\delta<\left | \lambda\right |.\label{31}
\end{equation}
The system (\ref{12})--(\ref{14}) with boundary conditions (\ref{16}), (\ref{17}), (\ref{26})--(\ref{30}) composes the eigenvalue problem for the parameter $\lambda$ for the investigations of a Townsend discharge stability.

\section{Diffusive instability of a Townsend discharge}

In this section, we solve system (\ref{12})--(\ref{14}) with boundary conditions (\ref{16}), (\ref{17})  in asymptotic limit $\delta=0$  and $\varepsilon_{e}\ll 1$, for which $\hat{E}=\hat{E}_{0}=1$. With nomenclature (\ref{22}) system of equation (\ref{12})--(\ref{14}) for for ion, $n^{(0)}_{i0}\left(z; \varepsilon_{e}\right)$, and electron, $n^{(0)}_{e0}\left(z; \varepsilon_{e}\right)$, densities  reduces to the following simple system
\begin{equation}
\frac{\partial n^{(0)}_{i0}\left(z; \varepsilon_{e}\right)}{\partial z}= -\hat{\alpha}n^{(0)}_{e0}\left(z; \varepsilon_{e}\right),\label{32}
\end{equation}
\begin{equation}
\varepsilon_{e}\frac{\partial^{2} n^{(0)}_{e0}\left(z; \varepsilon_{e}\right)}{\partial z^{2}}-\frac{\partial n^{(0)}_{e0}\left(z; \varepsilon_{e}\right)}{\partial z}+\hat{\alpha}n^{(0)}_{e0}\left(z; \varepsilon_{e}\right)=0.\label{33}
\end{equation}
The solution of Eqs.(\ref{32})--(\ref{33}) provides us with a Townsend discharge solution \cite{Amiranashvili}, extended on the accounting for the effects of electrons diffusion. With boundary conditions (see Eqs. (\ref{16}), (\ref{17}) and (\ref{26}, respectively)  $n^{(0)}_{i0}\left(z=1, \varepsilon_{e}\right)=0$, $n^{(0)}_{e0}\left(z=1, \varepsilon_{e}\right)=1$, $\partial n^{(0)}_{e0}\left(z=1, \varepsilon_{e}\right)/\partial z=0$,  the  expressions for electron, $n^{(0)}_{e0}\left(z, \varepsilon_{e}\right)$, and ion, $n^{(0)}_{i0}\left(z, \varepsilon_{e}\right)$, steady state densities are
\begin{eqnarray}
&\displaystyle n^{(0)}_{e0}\left(z, \varepsilon_{e}\right)=\frac{1}{\left(1-\frac{a_{1}}{a_{2}}\right)}
\nonumber \\
&\displaystyle \times \left(e^{-a_{1}\left( 1-z\right)}-\frac{a_{1}}{a_{2}}e^{-a_{2}\left( 1-z\right)}\right)\label{34}
\end{eqnarray}
\begin{eqnarray}
&\displaystyle n^{(0)}_{i0}\left(z, \varepsilon_{e}\right)=\frac{1}{\left(1-\frac{a_{1}}{a_{2}}\right)}
\nonumber \\
&\displaystyle \times
\frac{\alpha}{a_{1}}\left[\left(1-e^{-a_{1}\left( 1-z\right)}\right) -\frac{a_{1}^{2}}{a^{2}_{2}}\left(1-e^{-a_{2}\left( 1-z\right)}\right)\right] , \label{35}
\end{eqnarray}
where
\begin{eqnarray}
&\displaystyle a_{1,2}= \frac{1}{2\varepsilon_{e}}\pm\left(\frac{1}{4\varepsilon^
{2}_{e}}-\frac{\hat{\alpha}}{\varepsilon_{e}}\right)^{1/2} , \label{36}
\end{eqnarray}
The condition $n^{(0)}_{e0}\left(z=0, \varepsilon_{e}=0\right) =\gamma n^{(0)}_{i0}\left(z=0, \varepsilon_{e}=0\right)$ at cathode provides us with condition of the steady self-sustenance of Townsend discharge, extended on the accounting for the effect of electrons diffusion,
\begin{equation}
\gamma\frac{\alpha}{a_{2}}\left(e^{a_{2}}-1 \right)=1+\gamma\frac{\alpha}{a_{1}}\frac{a_{2}}{a_{1}}e^{a_{2}}, \label{37}
\end{equation}
which is, in fact, the equation which determines $E_{t}$, used in Eq.(\ref{8}) of our transformations to dimensionless variables. Last term in Eq.(\ref{37}) is negligibly small and it will be omitted in what follows. For $\varepsilon_{e}=0$ Eq.(\ref{37}) reduces to well known relation\cite{Raizer} $\gamma\left(e^{\hat{\alpha}}-1 \right)=1$. 

Using presentation (\ref{22}) with $\delta=0$, we obtain the system of equations for ion and electron densities perturbations,
\begin{eqnarray}
&\displaystyle \frac{\partial n^{(0)}_{i1}\left(z; \varepsilon_{e}\right)}{\partial z}-\lambda n^{(0)}_{i1}\left(z; \varepsilon_{e}\right)=-\hat{\alpha}n^{(0)}_{e1}\left(z; \varepsilon_{e}\right), \label{38}
\end{eqnarray}
\begin{eqnarray}
&\displaystyle -\varepsilon_{e}\frac{\partial^{2} n^{(0)}_{e1}\left(z; \varepsilon\right)}{\partial z^{2}}+\frac{\partial n^{(0)}_{e1}\left(z; \varepsilon_{e}\right)}{\partial z}=\hat{\alpha}n^{(0)}_{e1}\left(z; \varepsilon_{e}\right). \label{39}
\end{eqnarray}
The solution of that system with boundary conditions (\ref{17}) is
\begin{eqnarray}
&\displaystyle n^{(0)}_{e1}\left(z; \varepsilon_{e}\right)= C\left(e^{-a_{1}\left( 1-z\right)}-\frac{a_{1}}{a_{2}}
e^{-a_{2}\left( 1-z\right)} \right), \label{40}
\end{eqnarray}
and
\begin{eqnarray}
&\displaystyle n^{(0)}_{i1}\left(z; \varepsilon_{e}\right)= C\hat{\alpha}\left[e^{-\lambda\left( 1-z\right)}\left(\frac{1}
{a_{1}-\lambda}-\frac{a_{1}}{a_{2}}\frac{1}{a_{2}-\lambda} \right)\right.
\nonumber \\
&\displaystyle -\left.
\left(\frac{e^{-a_{1}\left( 1-z\right)}}{a_{1}-\lambda}-
\frac{a_{1}}{a_{2}}\frac{e^{-a_{2}\left( 1-z\right)}}{a_{2}-\lambda} \right)\right] , \label{41}
\end{eqnarray}
where $a_{1,2}$ are determined by Eq.(\ref{36}) and $C$ is arbitrary constant of the integration. By using solutions (\ref{39}) and (\ref{40}) in boundary condition (\ref{16}) we obtains the eigenvalue equation for parameter $\lambda$,
\begin{eqnarray}
&\displaystyle -\varepsilon_{e}a_{1}\left( e^{-a_{1}}-e^{-a_{2}}\right)+\left( e^{-a_{1}}-\frac{a_{1}}{a_{2}}e^{-a_{2}}\right)\nonumber  \\
&\displaystyle=\alpha\gamma\left[-\frac{1}{a_{1}-\lambda}\left(e^{-a_{1}}-e^{-\lambda} \right)
\right.
\nonumber \\
&\displaystyle +\left.\frac{a_{1}}{a_{2}}\frac{1}{a_{2}-\lambda}\left(e^{-a_{2}}-e^{-\lambda} \right) \right].\label{42}
\end{eqnarray}
Accounting for that $a_{1}\sim \varepsilon_{e}^{-1}\gg 1$, and $a_{1}\gg a_{2}\approx\hat{\alpha}\left( 1-\hat{\alpha}\varepsilon_{e}\right) $, Eq.(\ref{43}) may be simplified to
\begin{eqnarray}
&\displaystyle \left( \lambda-a_{2}\right)\left( 1-a_{2}\varepsilon_{e}\right)  =\alpha\gamma\left(1-e^{a_{2}-\lambda} \right), \label{43}
\end{eqnarray}
where we assume that $\lambda\neq a_{1}$ or $a_{2}$. For $\varepsilon_{e}=0$ Eq.(\ref{43}) has infinite number of complex  roots with negative real part, which corresponds to the damping oscillations of the ion and electron space charges perturbations. Also it has two evident real roots, $\lambda=a_{2}$ and $\lambda=0$. Root $\lambda=a_{2}$ is physically senseless, because for $\lambda =a_{2}$ ion density perturbation $n^{(0)}_{i1}\left(z; \varepsilon_{e}\right)$ becomes infinite.  $\lambda=0$ corresponds to stationary state, for which presentation (\ref{22}) becomes senseless. This root, however, due to the terms with finite $\varepsilon_{e}$ in Eq.(\ref{43}), will obtain relatively small non-zero value. Expanding the exponential in Eq.(\ref{43}) for small $\lambda$, we obtain simple solution for this root,
\begin{eqnarray}
&\displaystyle \lambda=\frac{a^{2}_{2}\varepsilon_{e}}{a_{2}+a_{2}\varepsilon_{e}-1-\hat{\alpha}\gamma}. \label{44}
\end{eqnarray}
Because the  denominator in Eq.(\ref{44}) is positive for the conditions of a Townsend discharge, root (\ref{43}) corresponds to the growth rate of the aperiodic instability, conditioned by electron diffusion. The growth rate (\ref{44}) is the main result of this paper. We identify the discovered instability as the diffusive instability, because the growth rate (\ref{44}) of this instability is proportional to the electron diffusion parameter $\varepsilon_{e}$. Because of the assumption $\delta =0$, the instability discovered is true Townsend discharge instability, which does not lead to the transition to other forms of glow discharge. In dimensional physical units, the growth rate (\ref{44}) is equal approximately to 
\begin{eqnarray}
&\displaystyle \lambda_{phys}\sim\frac{\alpha v_{s}^{2}}{L\nu_{in}}, \label{45}
\end{eqnarray}
where $v_{s}^{2}=T_{e}/m_{i}$ is the ion sound velocity, $\nu_{in}$ is ion-neutral collision frequency, and Einstein relation, $D_{e}=T_{e}\mu_{e}/e$, was used. For a discharge in Helium under the pressure of 30 Torr, with $L=1$ mm distance between the electrodes, and electron temperature $T_{e}=1$ eV, the growth rate (\ref{45}) is equal approximately to $\lambda_{phys}\backsimeq 2\cdot 10^{4}$ c$^{-1}$.

\section{Conclusion}

In this work, by solving the eigenvalue problem for the parameter $\lambda$, we have shown, that a Townsend discharge is unstable due to a joint action of the ionization process and electron diffusion. Discovered instability is responsible for the spatially homogeneous exponential growth with time the perturbations of the electron and ion densities due to the ionization processes and from this point of view it may be considered as a kind of the ionization instability\cite{Raizer}. However it is principally different from known ionization instability of a Townsend discharge\cite{Kaganovich, Kolobov}.  The dispersion equation (\ref{42}) is derived from the boundary value problem solution for which the electron diffusion have to be included in the boundary conditions (and which can be neglected for the main part of the gap between the electrodes). The growth rate (\ref{44}) is proportional to the ionisation parameter $\hat{\alpha}$ $\left(a_{2}\simeq  \hat{\alpha} \right)$ and does not depend on the local properties of a Townsend discharge with practically homogeneous electric field, not disturbed by spatial charge, whereas the well known ionization instability of a Townsend discharge is developed due to violation of the ionization balance resulted from the local spontaneous fluctuation of the spatial charge. Accounting for this and that the growth rate (\ref{44}) is proportional to the electron diffusion parameter $\varepsilon_{e}$, we identify the discovered instability as the diffusive instability. The presented theory is valid for sufficiently small currents and /or strong applied electric field $E_{t}$, for which condition (\ref{31}) is valid.  However, because of the gradual growth of the ion and electron densities with time and concomitant to this process growth of the discharge current, that in turn leads to the growth of the parameter $\delta$, the effects of the small space charge, $\delta \ll 1$, have to be addressed on following stages of the discharge development. Therefore, discovered diffusive instability may be considered as a precursor of the ordinary spatially inhomogeneous ionization instability, as well as an unstable background for its development at times $\lambda_{phys}^{-1}$.

\begin{acknowledgments}
The first author would like to thank Prof. M.S.~Benilov for valuable discussions and encouragement. 
The work was partially supported by project PTDC/FIS/68609/2006 of FCT, Centro de Cincias Matemticas of FCT, Portugal, and Pusan National University, Republic of Korea.
\end{acknowledgments}

\end{document}